\documentclass[cits]{PoS}
\pdfoutput=1
\def\apjl{ApJ}%
\def\mnras{MNRAS}%
\def\prd{Phys.~Rev.~D}%
\def\prl{Phys.~Rev.~Lett.}%

\title{Impacts of WIMP dark matter upon stellar evolution: main-sequence stars}

\ShortTitle{WIMPs and main-sequence stellar evolution at the Galactic Centre}

\author{\speaker{Pat Scott}\\
        Cosmology, Particle Astrophysics and String Theory, Physics, Stockholm University and\\ 
	Oskar Klein Centre for Cosmoparticle Physics\\
	AlbaNova University Centre, SE-106 91 Stockholm, Sweden\\
        E-mail: \email{pat@fysik.su.se}}

\author{Malcolm Fairbairn\\
	Theory Division, CERN, CH-1211, Geneva 23, Switzerland and \\
	Physics, Kings College London, Strand, London WC2R 2LS, UK\\
        E-mail: \email{malc@cern.ch}}

\author{Joakim Edsj\"o\\
        Cosmology, Particle Astrophysics and String Theory, Physics, Stockholm University and\\ 
	Oskar Klein Centre for Cosmoparticle Physics\\
	AlbaNova University Centre, SE-106 91 Stockholm, Sweden\\
        E-mail: \email{edsjo@fysik.su.se}}

\abstract{The presence of large amounts of WIMP dark matter in stellar cores has been shown to have significant effects upon models of stellar evolution.  We present a series of detailed grids of WIMP-influenced stellar models for main sequence stars, computed using the \textsf{DarkStars} code.  We describe the changes in stellar structure and main sequence evolution which occur for masses ranging from 0.3 to 2.0 M$_\odot$ and metallicities from $Z=0.0003$--$0.02$, as a function of the rate of energy injection by WIMPs.  We then go on to show what rates of energy injection can be obtained using realistic orbital parameters for stars near supermassive black holes, including detailed considerations of dark matter halo velocity and density profiles.  Capture and annihilation rates are strongly boosted when stars follow elliptical rather than circular orbits, causing WIMP annihilation to provide up to 100 times the energy of hydrogen fusion in stars at the Galactic centre.}

\FullConference{Identification of dark matter 2008\\
		 August 18-22, 2008\\
		 Stockholm, Sweden}

\begin{document}

The impacts of weakly-interacting massive particles (WIMPs) upon stellar structure and evolution are emerging as a promising means for the indirect detection of dark matter.  The most interesting theoretical results have been found at the Galactic centre (GC) \cite{GC1,GC2} and in the early universe \cite{PopIII}.  At the GC, large number of WIMPs could become gravitationally bound to stars through weak scattering events.  These bound WIMPs would then undergo further collisions with atomic nuclei and sink to the centres of stars as they lose progressively more energy.  If the WIMPs were Majorana particles (as most well-motivated WIMP candidates tend to be), they would then self-annihilate.  For typical stellar core densities and particle physics scenarios, roughly 90\% of the rest-mass energy of each WIMP would ultimately be injected into the core as heat; the remaining fraction would be lost as neutrinos.

We have developed a highly-detailed `dark' stellar evolution code (\textsf{DarkStars}) in order to simulate capture and annihilation of WIMPs, and the subsequent impacts upon stellar evolution.  Following on from our initial work based on a preliminary version of this code \cite{GC2}, 
\begin{figure}[bh]
\begin{minipage}[t]{0.48\textwidth}
\includegraphics[width=\linewidth, trim = 0 0 0 30, clip=true]{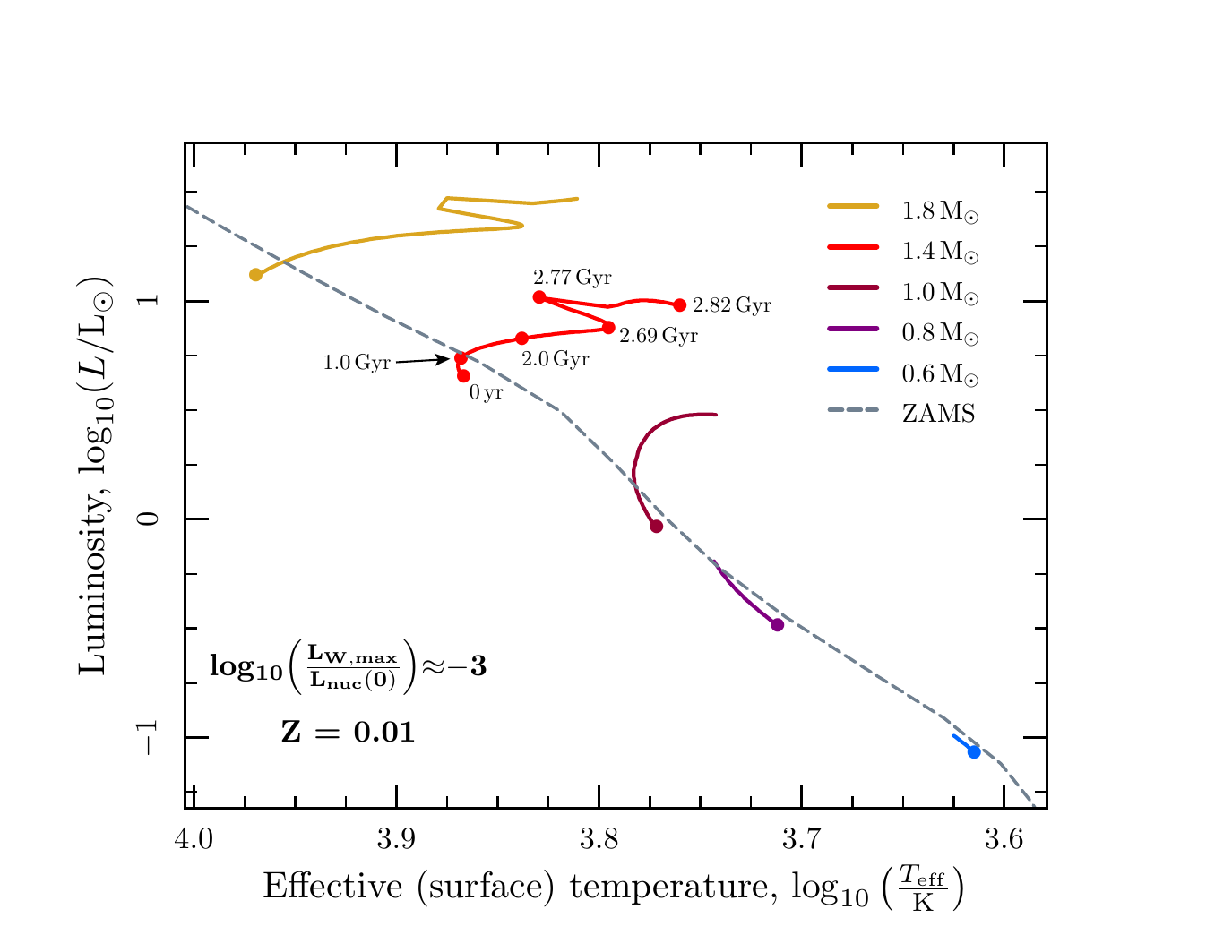}
\includegraphics[width=\linewidth, trim = 0 0 0 30, clip=true]{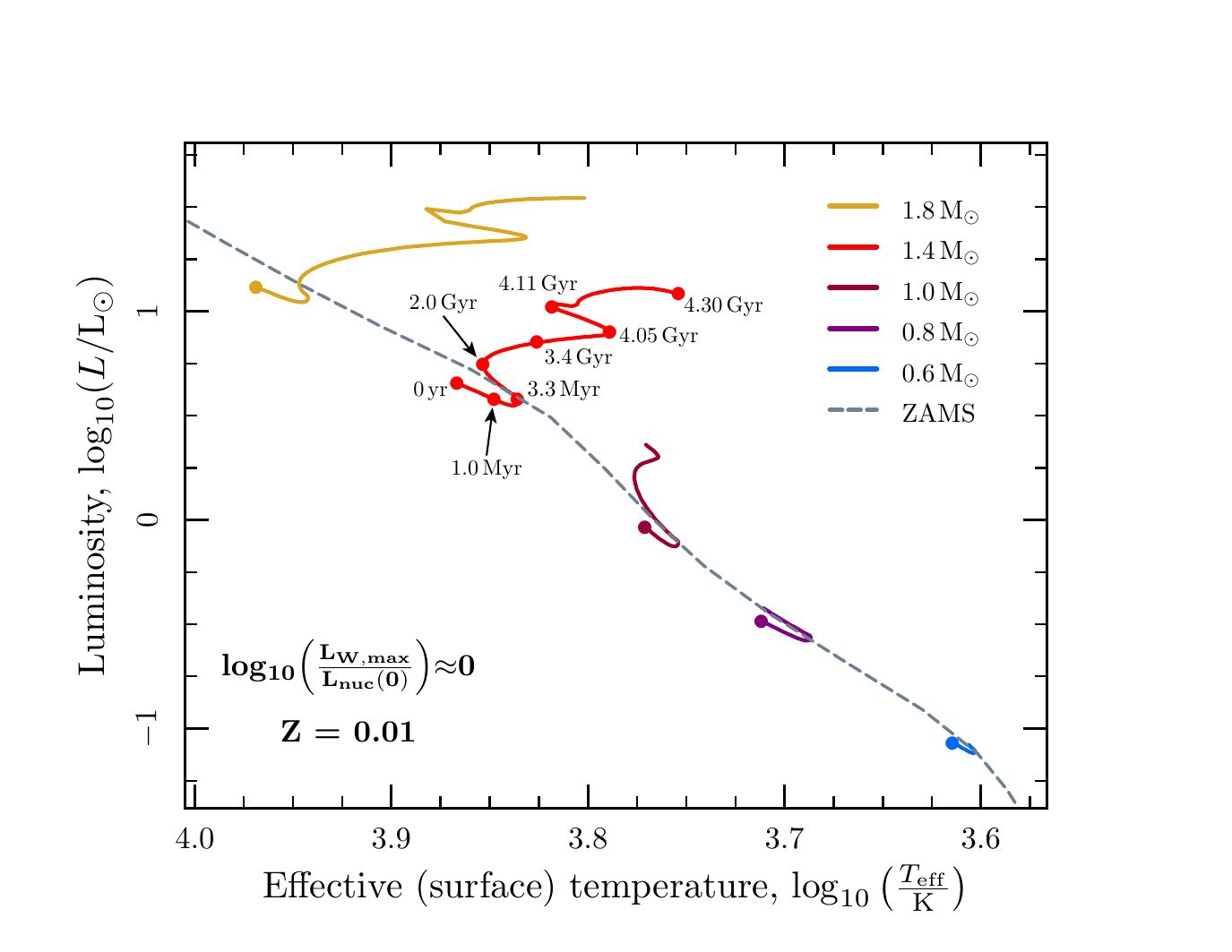}
\end{minipage}
\hspace{0.04\textwidth}
\begin{minipage}[t]{0.48\textwidth}
\includegraphics[width=\linewidth, trim = 0 0 0 30, clip=true]{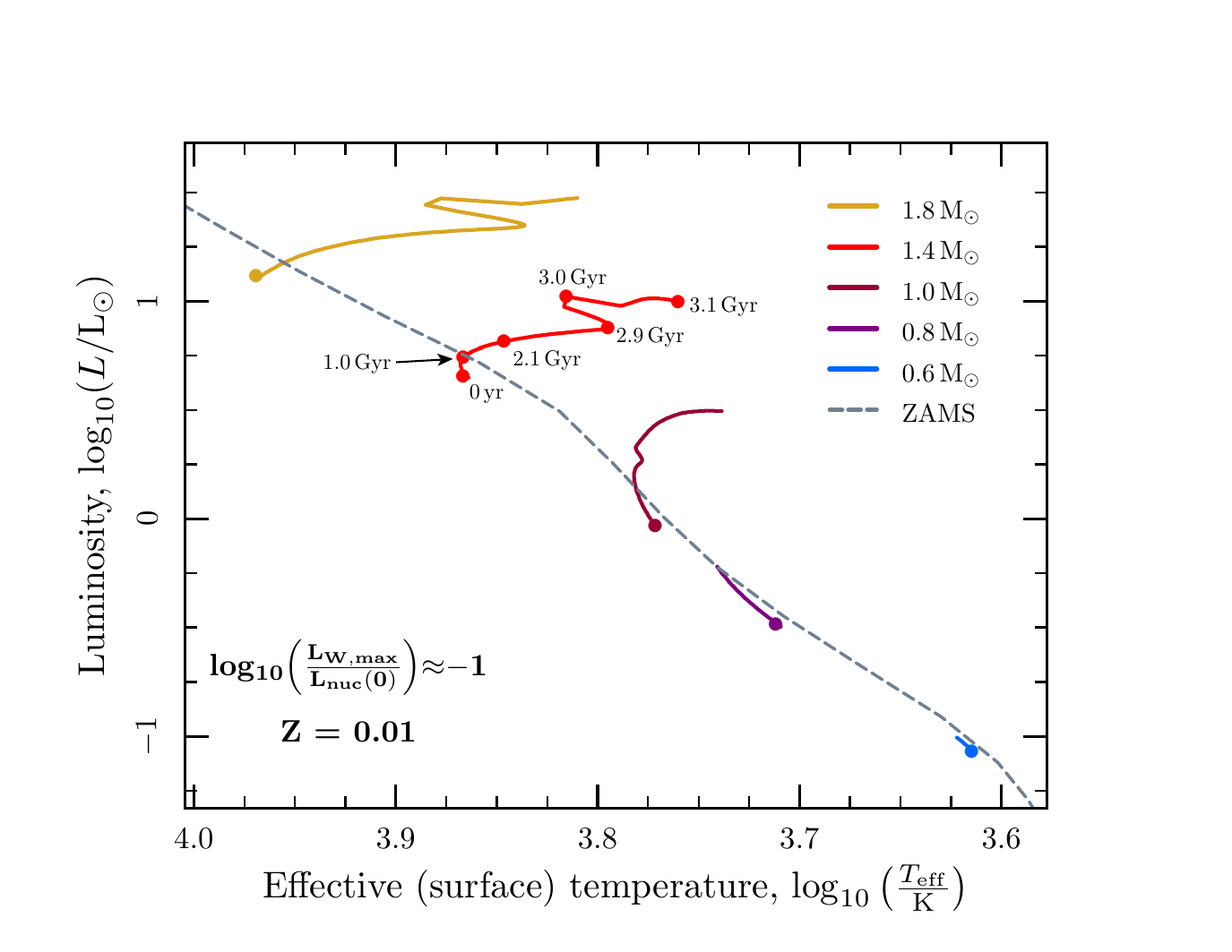}
\includegraphics[width=\linewidth, trim = 0 0 0 30, clip=true]{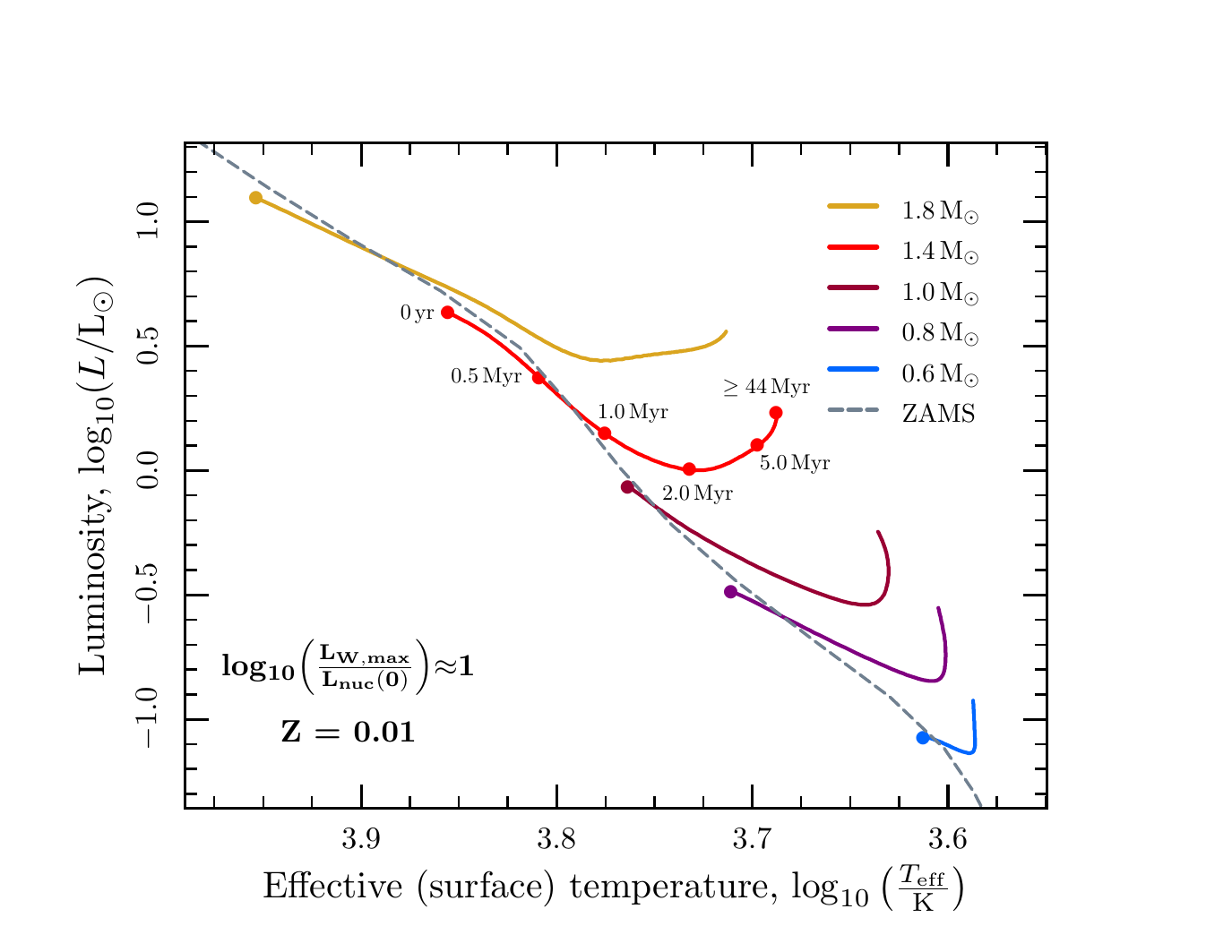}
\end{minipage}
\caption{HR diagram showing the evolutionary tracks followed by stars of different masses, as WIMPs are allowed to provide different fractions of the stars' total energy budgets.  The energy provided by WIMP annihilation is shown in the bottom left of each sub-plot as the ratio of the maximum luminosity achieved by WIMP annihilation to the initial luminosity due to fusion.  Starting points of tracks are indicated with filled, unlabelled circles, whilst labelled circles give indicative ages during the evolution of 1.4\,M$_\odot$ stars.  Simulations have been halted when the star exhausts its core hydrogen supply or reaches the current age of the universe.  Stars with a greater luminosity contribution from WIMPs push further up the Hayashi track and spend longer there before returning to the main sequence.  Those which come to be entirely dominated by WIMP annihilation (\emph{bottom right}) evolve back up the Hayashi track on the thermal timescale and halt, holding their position well beyond the age of the universe.}
\label{Fig1}
\end{figure}
we have recently presented an in-depth investigation of the evolutionary impacts of WIMP capture and annihilation upon main-sequence stars at the GC \cite{Scott08b}.  Here we summarise this work; the reader is referred to the full-length paper for extensive discussions of existing literature, the code, input physics, results, and prospects for detection.

\begin{figure}
\begin{center}
\includegraphics[width=0.7\columnwidth, trim = 0 0 0 30, clip=true]{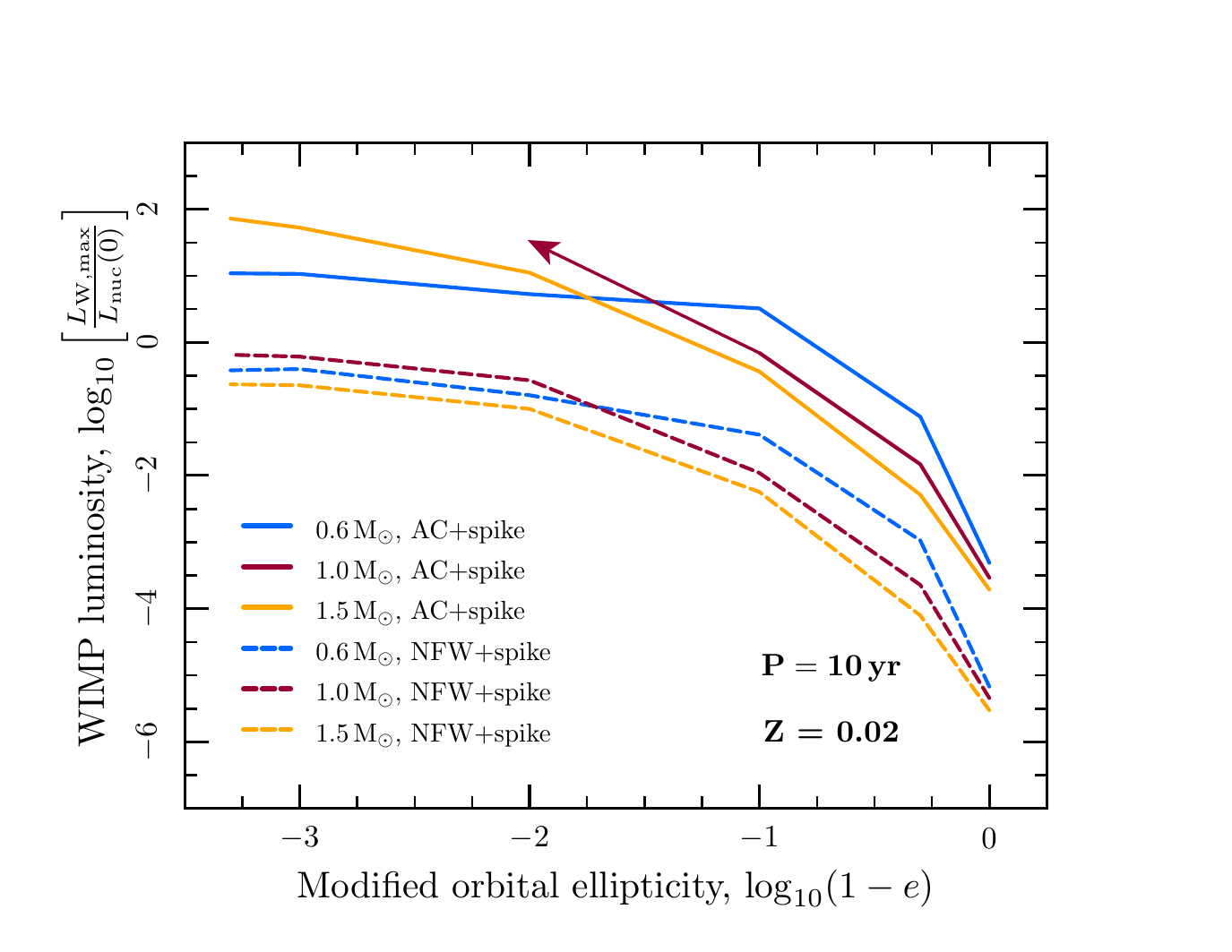}
\end{center}
\caption{WIMP luminosities achievable by stars on 10-year orbits about the GC.  On realistic orbits, annihilation can provide up to 100 times the power of nuclear fusion.  If the Galactic halo has undergone adiabatic contraction (AC+spike), break-even between fusion and annihilation energy occurs in stars on any orbit with an eccentricity $e\gtrsim 0.9$, for all masses $M_\star\lesssim1.5$\,M$_\odot$.  If not (NFW+spike), annihilation begins to rival nuclear fusion in stars of a solar mass or less on orbits with $e\gtrsim0.99$.  These curves have been obtained by calculating the capture rates on such orbits near the GC, applying small boosts due to the non-Gaussian distribution of WIMP velocities, and converting to the maximum WIMP-to-nuclear burning ratios expected during a star's evolution (see \protect\cite{Scott08b} for details).  The arrow indicates that the 1\,M$_\odot$, AC+spike curve is expected to continue upwards, but there is no reliable way to convert capture rates to WIMP luminosities in this region because it is beyond the range of parameters we considered in our grid of stellar evolutionary models.}
\label{Fig2}
\end{figure}

We evolved a grid of model stars immersed in dark matter halos of various concentrations, with stellar masses ranging from 0.3 to 2.0 M$_\odot$ and metallicities from $Z=0.0003$--$0.02$.  As the external WIMP concentration is raised, WIMP annihilation provides a larger fraction of the star's total luminosity.  For any given external WIMP density, this fraction is higher for lower mass stars.  For large enough WIMP luminosities, stars evolve back up the protostellar Hayashi track on the thermal timescale (Fig.~\ref{Fig1}).  The higher the WIMP luminosity, the further a star will retreat up the Hayashi track, and the longer it will pause there before returning to the main sequence.  For very high WIMP densities, the duration of this phase can greatly exceed the present age of the Universe.

Stellar orbits very close to the GC are expected to result in high WIMP capture rates due to the high density of dark matter there.  They are also expected to result in very high orbital velocities about the central supermassive black hole, owing to its large mass.  This acts to reduce capture rates, as it means a very high mean star-WIMP relative velocity, which reduces the probability that any given scattering event will reduce the total energy of a WIMP enough for it to become gravitationally bound to a star.  We find that stars moving on circular orbits, where the orbital velocity is constant, can never capture sufficient WIMPs for their structure or evolution to be altered; this is true of any circular orbit in the Milky Way.  Should they move on elliptical orbits instead, stars can slow quite significantly relative to the dark matter halo when they reach apoapsis.  This has the effect of drastically boosting capture rates (Fig.~\ref{Fig2}).  

Depending upon the halo model employed, we find that low-mass stars on sufficiently elliptical orbits can achieve annihilation luminosities of up to 100 times their fusion luminosities.  Should such stars be observed at the GC, they would constitute unequivocal evidence for WIMP dark matter.  Were normal stars instead observed on the same orbits, one could place stringent limits on the mass, spin-dependent nuclear-scattering cross-section, halo density and velocity distribution of WIMPs at the GC.  Presently only high-mass stars have been observed on such tight elliptical orbits, but as observations improve over the next few years, any low-mass counterpart population should become observationally accessible.

\end{document}